\documentclass[prb,twocolumn,amsmath,amssymb]{revtex4}
\begin{document}
\author{Kevin Leung$^{1*}$ and Joanne L.~Budzien$^2$}
\affiliation{$^1$Surface and Interface Sciences Department, MS 1415,
Sandia National Laboratories, Albuquerque, New Mexico 87185, USA; 
{\tt kleung@sandia.gov} \\
$^2$Department of Physics and Engineering 
Frostburg State University, Frostburg, MD 21532, USA}
\date{\today}
\title{{\it Ab initio} Molecular Dynamics Simulations of the Initial
Stages of Solid-electrolyte Interphase Formation on Lithium Ion Battery
Graphitic Anodes}

\input epsf
 
\begin{abstract}
 
The decomposition of ethylene carbonate (EC) during the initial growth
of solid-electrolyte interphase (SEI) films at the solvent-graphitic
anode interface is critical to lithium ion battery operations.
{\it Ab initio} molecular dynamics simulations of explicit liquid
EC/graphite interfaces are conducted to study these electrochemical
reactions.  We show that carbon edge terminations are crucial
at this stage,
and that achievable experimental conditions can lead to surprisingly fast EC
breakdown mechanisms, yielding decomposition products seen in experiments
but not previously predicted.
\end{abstract}
 
\maketitle
 
\begin{figure}
\centerline{\hbox{\epsfxsize=3.00in \epsfbox{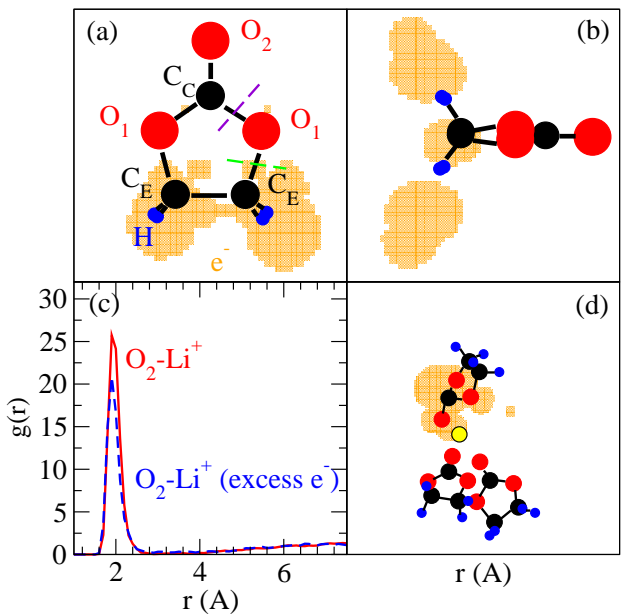}} }
\caption[]
{\label{fig1} \noindent
Black, red, blue, and yellow spheres: C, O, H, and Li.
Orange shading: regions of spin (i.e., excess
$e^-$) density $\rho_s > 8$$\times$10$^{-3}$~$|e|$\AA$^{-3}$.
(a) \& (b) Isolated EC$^-$.  
Green and violet lines depict two breakdown modes; the latter is
observed only with explicit treatment of liquid EC (see text).
(c) $g(r)$ between Li$^+$ and EC O$_2$ site for systems with and without
excess $e^-$.  (d) Snapshot of Li$^+$ coordinated to intact EC$^-$ and
two other EC molecules (other EC in the liquid omitted).
}
\end{figure}

Improving the fundamental scientific understanding of lithium ion
batteries\cite{book2,book,review} is critical for electric vehicles and
efficient use of solar and wind energy.  A key limitation in current batteries
is their reliance on passivating solid electrolyte interphase (SEI) films
on graphitic anode surfaces.\cite{book2,book,review,intro1,intro2}  Upon first
charging of a pristine battery, the large negative potential applied to induce
Li$^+$ intercalation into graphite decomposes ethylene carbonate (EC,
Fig.~\ref{fig1}) molecules in the solvent, yielding a self-limiting,
30-50~nm thick, passivating SEI layer containing Li$_2$CO$_3$,
lithium ethylene dicarbonate
((CH$_2$CO$_3$Li)$_2$),\cite{book,intro1,intro2,ethbicarb} and salt
decomposition products.  C$_2$H$_4$ and CO gases have also been
detected\cite{yoshida,ogumi} and shown to come from EC.\cite{c13}
Similar reactions occur during power cycling when the SEI film cracks
and graphite is again exposed to EC.\cite{book}  If instead the solvent is
pure propylene carbonate (PC), a stable SEI film does not
materialize\cite{book2,book} and the battery fails.  Our work shows
that novel mechanisms for the initial stages of SEI-growth at 
electrode-electrolyte interfaces can be simulated
within time scales accessible to {\it ab initio} molecular dynamics
(AIMD),\cite{cpmd} which have successfully modelled liquid-solid
interfaces.\cite{surface_aimd}  AIMD is likely also applicable
to shed light on cosolvent/additives which must decompose more readily
than EC to alter and improve SEI structure, Li$^+$ transport, and
passivating properties.\cite{book2,book}
 
EC-decomposition mechanisms under electron-rich conditions
have been proposed (e.g., Refs.~\onlinecite{intro1,intro2})
and investigated using gas cluster Density Functional
Theory calculations with and without dielectric continuum
approximation of the liquid environment.\cite{bal01,bal02,vollmer,han}
Thus ``EC$^-$'', coordinated to Li$^+$ or otherwise, has been
predicted to undergo ethylene carbon (C$_{\rm E}$)-oxygen (O$_1$) bond
cleavage to form a more stable radical anion (Figs. 1a-b).  The
potential energy barrier involved is at least 0.33~eV.\cite{bal01,han}
Carbonyl carbon (C$_{\rm C}$)-O$_1$ bond-``breaking''
(or elongation) in the gas phase EC$^-$-Li$^+$ complex yields a
lower barrier, but metastable products.\cite{han}
 
Unlike these previous work, AIMD simulations can include explicit
liquid state environments and EC/graphite interfaces.  Unlike 
classical force field-based simulations,\cite{borodin,li_balbuena}
AIMD accounts for covalent bond-breaking.
We apply the VASP code,\cite{vasp,vasp1} the Perdew-Burke-Ernzerhof (PBE)
functional,\cite{pbe} $\Gamma$-point Brillouin zone sampling,
400~eV planewave energy cutoff, tritium masses for all protons to allow
Born-Oppenheimer dynamics time steps of 1~fs, and a 10$^{-6}$~eV
energy convergence criterion.  The hybrid PBE0 functional,\cite{pbe0,pbe0a}
more accurate in many cases, is used for spot checks.
A Nose thermostat maintains the temperature 
at T=450~K to avoid EC freezing.  A uniform neutralizing background charge
is imposed on systems with net charges.\cite{init}
Non-adiabatic quantum effects,\cite{rossky1} not generally evoked in
EC-breakdown reactions,\cite{bal01,bal02,han,vollmer} are neglected
and will be examined in the future.  Some gas phase calculations are also
conducted using the Gaussian suite of programs and a 6-311++G(d,p)
basis.\cite{gauss}
 
We first consider liquid EC to motivate subsequent breakdown
products at the EC-graphite interface.  A 17~ps trajectory with
32~EC molecules and a Li$^+$, but no excess $e^-$, is conducted
in a (15.24~\AA)$^3$ simulation cell, corresponding to the experimental
1.32~g/cc EC density.\cite{li_balbuena}  Integrating
over the first peak of the pair correlation function ($g(r)$, Fig.~\ref{fig1}c)
between Li$^+$ and O atoms yields a solvation number $N_{\rm EC}=4.0$,  in
agreement with a classical force field prediction\cite{li_balbuena,new}
but slightly smaller than the Raman value of 4.9.\cite{hyodo}
 
Next we add an excess $e^-$ to three randomly chosen configurations
along the Li$^+$-32~EC trajectory and restart AIMD.  (Liquid state
chemical reactions require multiple initial conditions to
account for the disorder.)  The 9.7~kcal/mol C$_{\rm E}$-O$_1$
cleavage barrier previously predicted\cite{bal01} implies 10$^{-9}$s
lifetimes at T=450~K if one assumes a vibrational prefactor of
10$^{-13}$s$^{-1}$.  The $g(r)$ from one 20~ps AIMD run is indeed
consistent with this prediction (Fig.~\ref{fig1}c).  An excess electron
is trapped on an EC near the Li$^+$.  Its $N_{\rm EC}$ averages
to 3.3; no bond breaking, even of the C$_{\rm C}$-O$_1$ type,\cite{han}
is observed.  An AIMD snapshot confirms that the excess
$e^-$ is trapped on a EC$^-$-Li$^+$ complex,\cite{bal01} at the
carbonate end (Fig.~\ref{fig1}d).
 
\begin{figure}
\centerline{(a) \hbox{\epsfxsize=1.48in \epsfbox{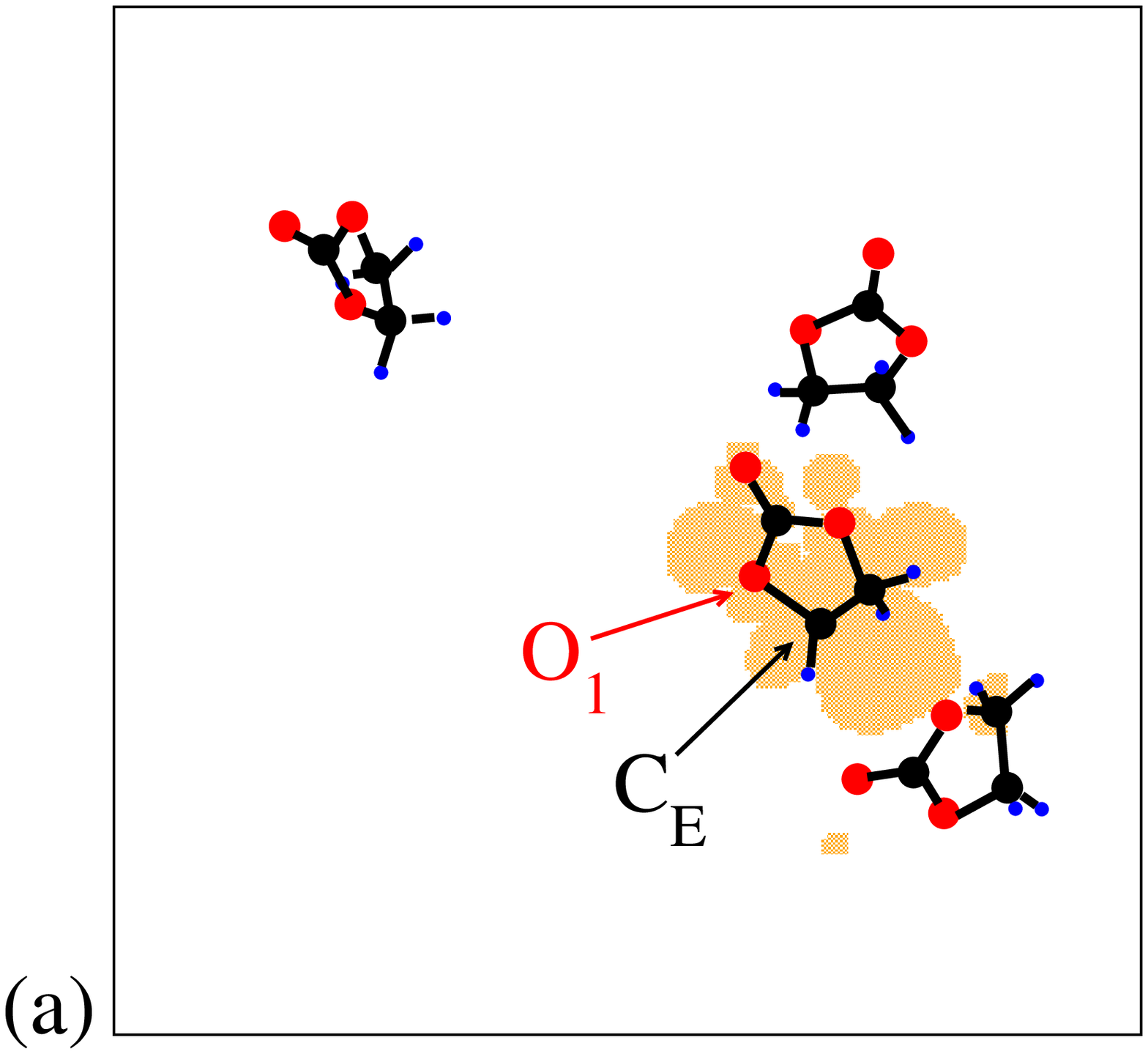}}
            \hbox{\epsfxsize=1.50in \epsfbox{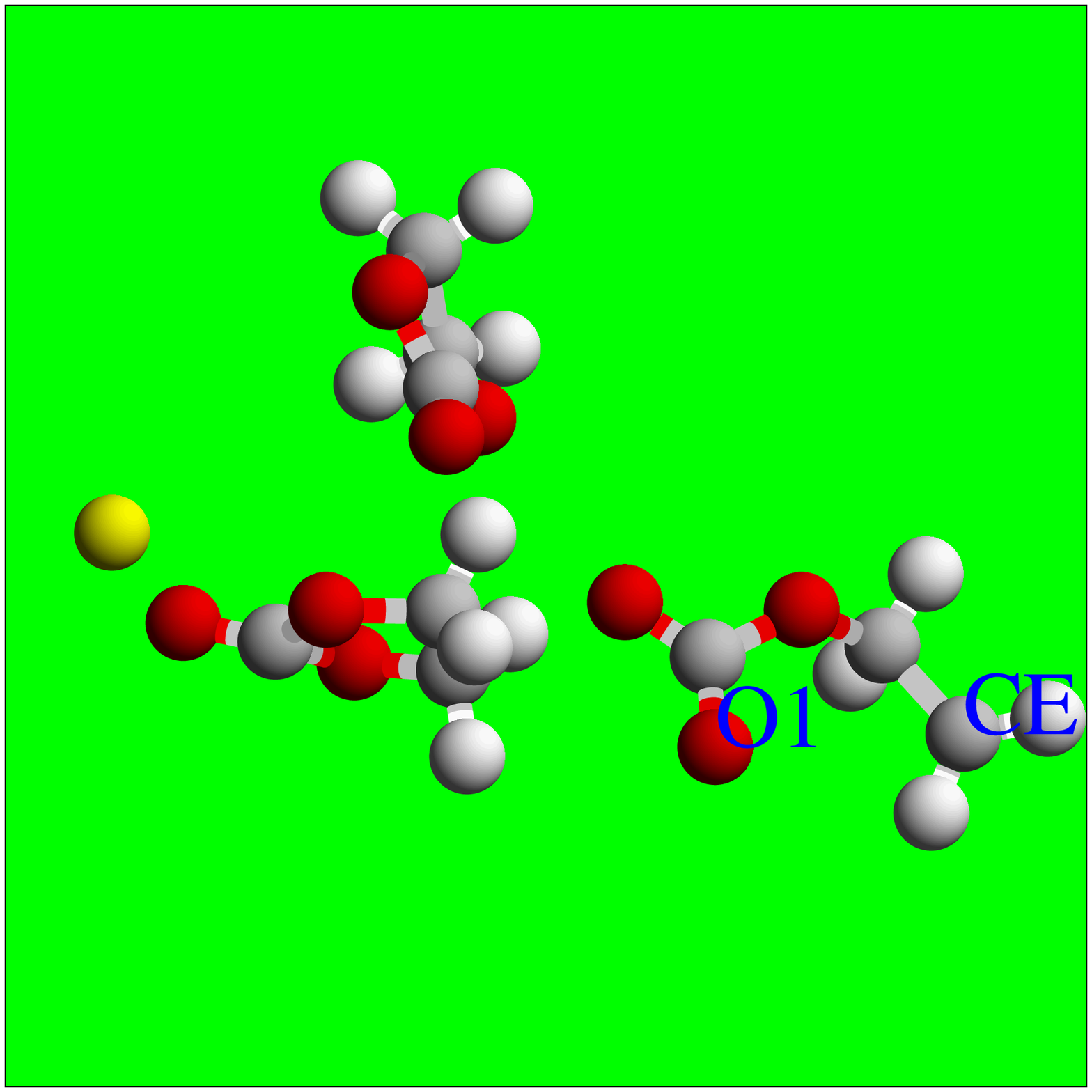}} (b)}
\centerline{(c) \hbox{\epsfxsize=1.50in \epsfbox{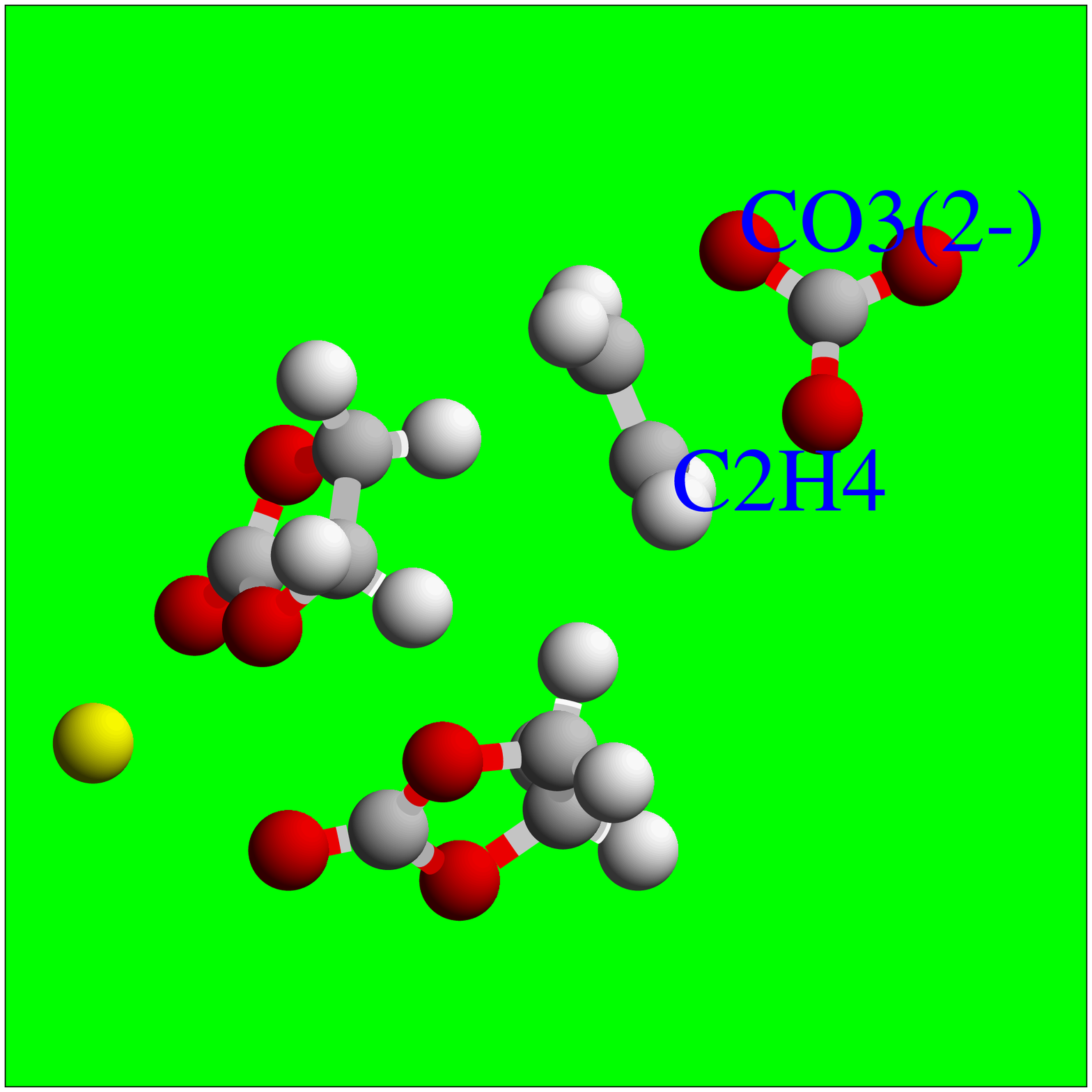}}
            \hbox{\epsfxsize=1.50in \epsfbox{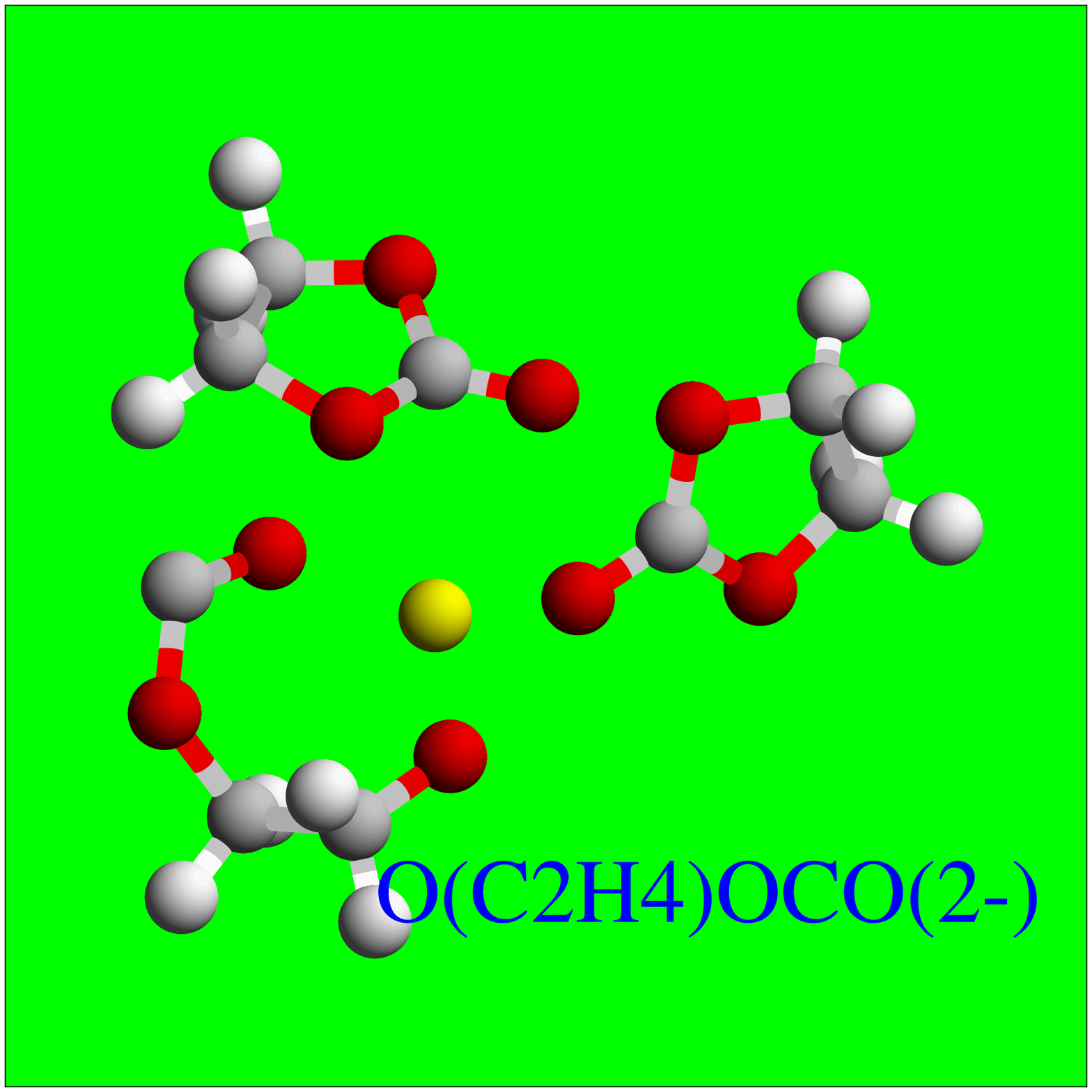}} (d)}
\caption[]
{\label{fig2} \noindent
(a) In this snapshot, substantial $e^-$ density resides on the EC with
a C$_{\rm E}$-O$_1$ bond about to break.
(b) C$_E$-O$_1$ bond breaks in the EC$^-$ soon afterwards.
(c) \& (d) EC$^-$ + $e^-$ $\rightarrow$
C$_2$H$_4$ +  CO$_3^{2-}$ or O(C$_2$H$_4$)OCO$^{2-}$, after
50~fs and 1.0~ps respectively.  
Only a few select EC are shown;
the colors are as in Fig.~\ref{fig1} except that C and H are grey
and white in panels (b)-(d).
}
\end{figure}

In contrast, two other starting configurations lead to a new mechanism:
a C$_{\rm E}$-O$_1$ bond breaks irreversibly in two distinct EC molecules 
outside the Li$^+$ first solvation shell within a
surprisingly short 0.8~ps (Fig.~\ref{fig2}a-b).  Here the excess
$e^-$ may have avoided the Li$^+$ vicinity because the Li$^+$ N$_{\rm EC}$
may need to spontaneously decrease to accommodate an electron
(Fig.~\ref{fig1}c), leading to an kinetic barrier.  This mechanism may
dominate in high excess $e^-$ concentration ([$e^-$]) regions near the
anode if most Li$^+$ there have already
complexed with EC$^-$.  While the PBE functional we use might slightly
underestimate barriers compared to hybrid functionals,\cite{bal01,han}
it still predicts a barrier of 0.37~eV for isolated EC$^-$
and a free energy barrier of 0.33~eV when a dielectric continuum solvation
is added. (See the electronic supporting information, ESI.)
Furthermore, gas phase EC$^-$ is not observed to crack
within a 7~ps AIMD trajectory.  Thus explicit treatment
of solvent EC molecules, which have large dipole moments,
appears responsible for the fast reaction rate.  Given the
use of the tritium masses and the slightly lower PBE functional reaction
barriers, we should focus on relative (not absolute) time scales of different
mechanisms, not absolute values.
 
The disparity between gas and solution phase rates likely arises
from the repulsive electron affinity of gas phase EC.\cite{bal01}  While
electronic structure calculations can impose a metastable isolated
``EC$^-$,'' the excess $e^-$ lies
{\it outside} the molecule (Fig.~\ref{fig1}b);
%\cite{note1} 
the C$_{\rm E}$-O$_1$ anti-bonding orbital is not occupied and
bond-breaking is not facilitated.  In contrast, AIMD/PBE simulations
predict that, in liquid EC, the excess $e^-$ is initially delocalized
{\it within} one or more EC molecules, unlike in water where excess
$e^-$ occupies intermolecular spaces stabilized by hydrogen-bond
donors.\cite{rossky1}  When instantaneous favorable molecular
geometries localize the $e^-$ on one EC and substantially populates 
orbitals on C and O atoms (Fig.~\ref{fig2}a), bond-breaking pathways
with rates different from that in the gas phase emerge.
Using a EC liquid snapshot, we confirm that the 
spatial distribution of excess $e^-$ computed using PBE and PBE0
functionals are qualitatively similar.\cite{new} 
This behavior depends on the LUMO (lowest unoccupied molecular orbital)
level; while PBE and PBE0 predicts gas phase HOMO (highest occupied
MO)-LUMO gaps of 6~eV and 8.2~eV respectively, the PBE0 LUMO energy level
is higher by only 0.45~eV.  
 
Doubling [$e^-$] by adding another spin-antiparallel $e^-$
to the Fig.~\ref{fig2}b liquid configuration yields
a C$_2$H$_4$/CO$_3^{2-}$ pair within 50~fs (Fig.~\ref{fig2}c).
%This is consistent with the high reduction potential
%predicted.\cite{vollmer}
Starting with the EC$^-$-Li$^+$ complex in Fig.~\ref{fig1}d
and adding an $e^-$ leads to C$_C$-O$_1$ bond cleavage 
and a O(C$_2$H$_4$)OCO$^{2-}$ instead (Fig.~\ref{fig2}d; see
ESI for charge state analyses).  This bond was previously shown to be
the weaker bond when even one electron is added to EC.\cite{han}
The latter fast 2-$e^-$ mechanism
is not seen without explicit solvent.\cite{vollmer}  

\begin{figure}
\centerline{(a) \hbox{\epsfxsize=1.50in \epsfbox{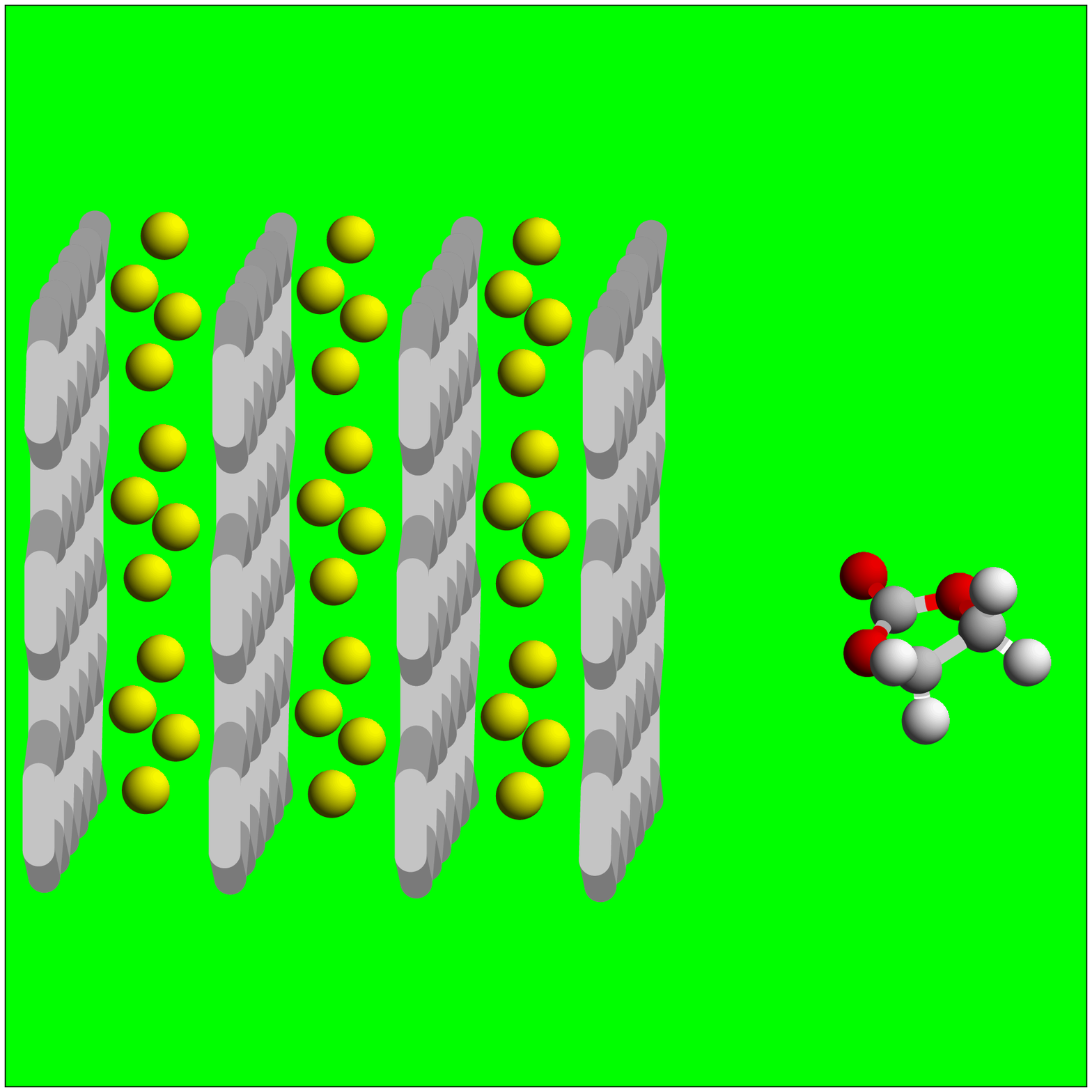}}
            \hbox{\epsfxsize=1.50in \epsfbox{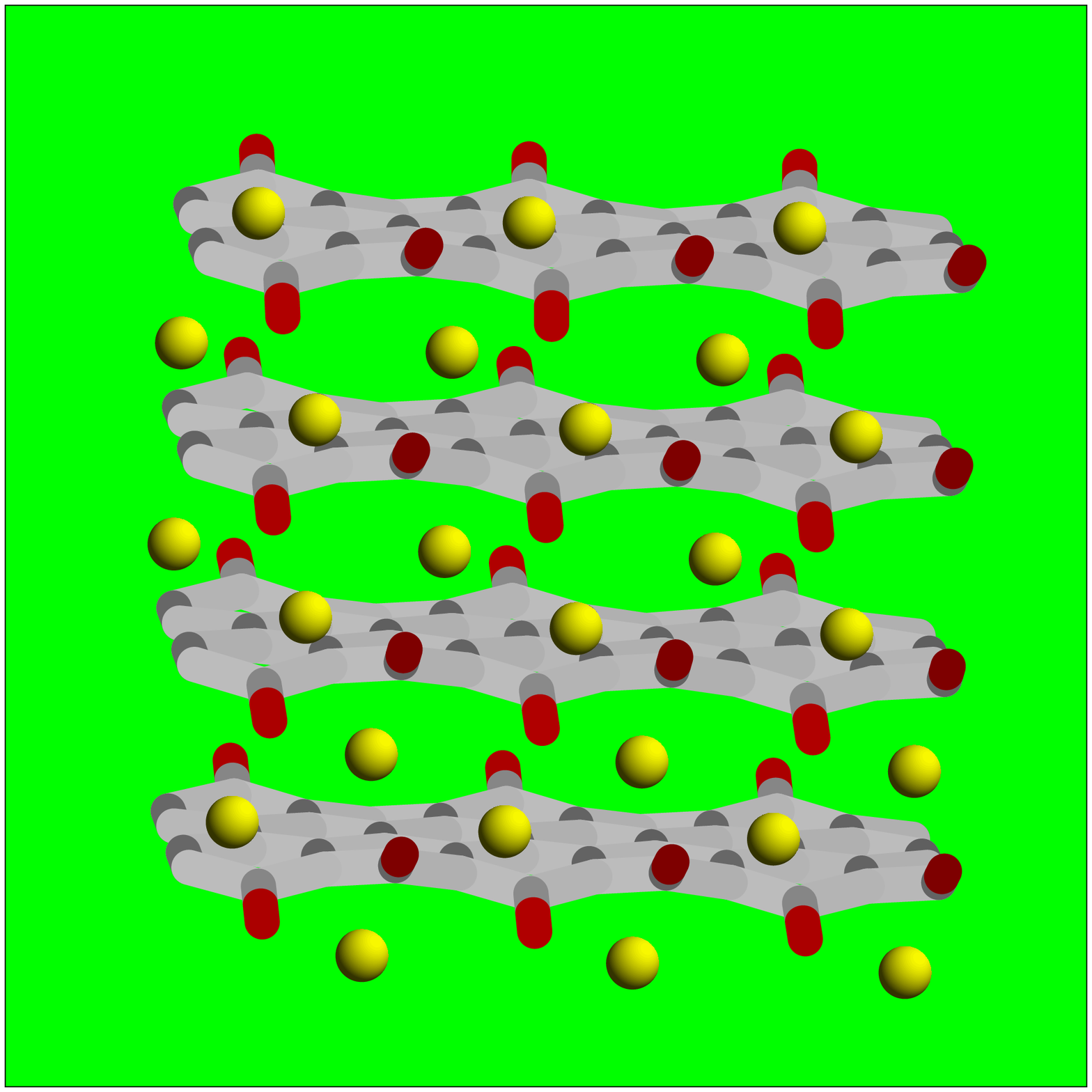}} (b)}
\centerline{(c) \hbox{\epsfxsize=1.50in \epsfbox{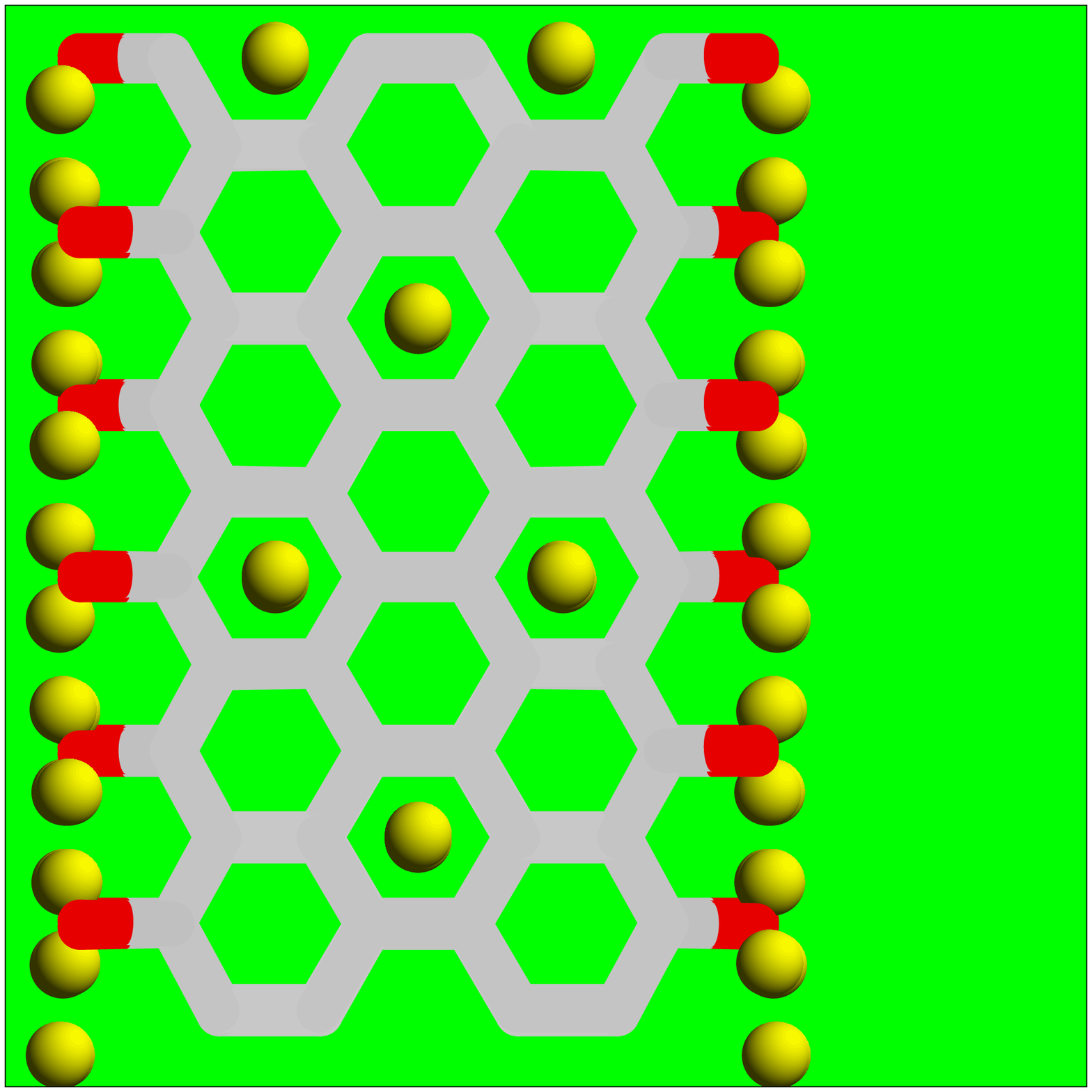}}
            \hbox{\epsfxsize=1.50in \epsfbox{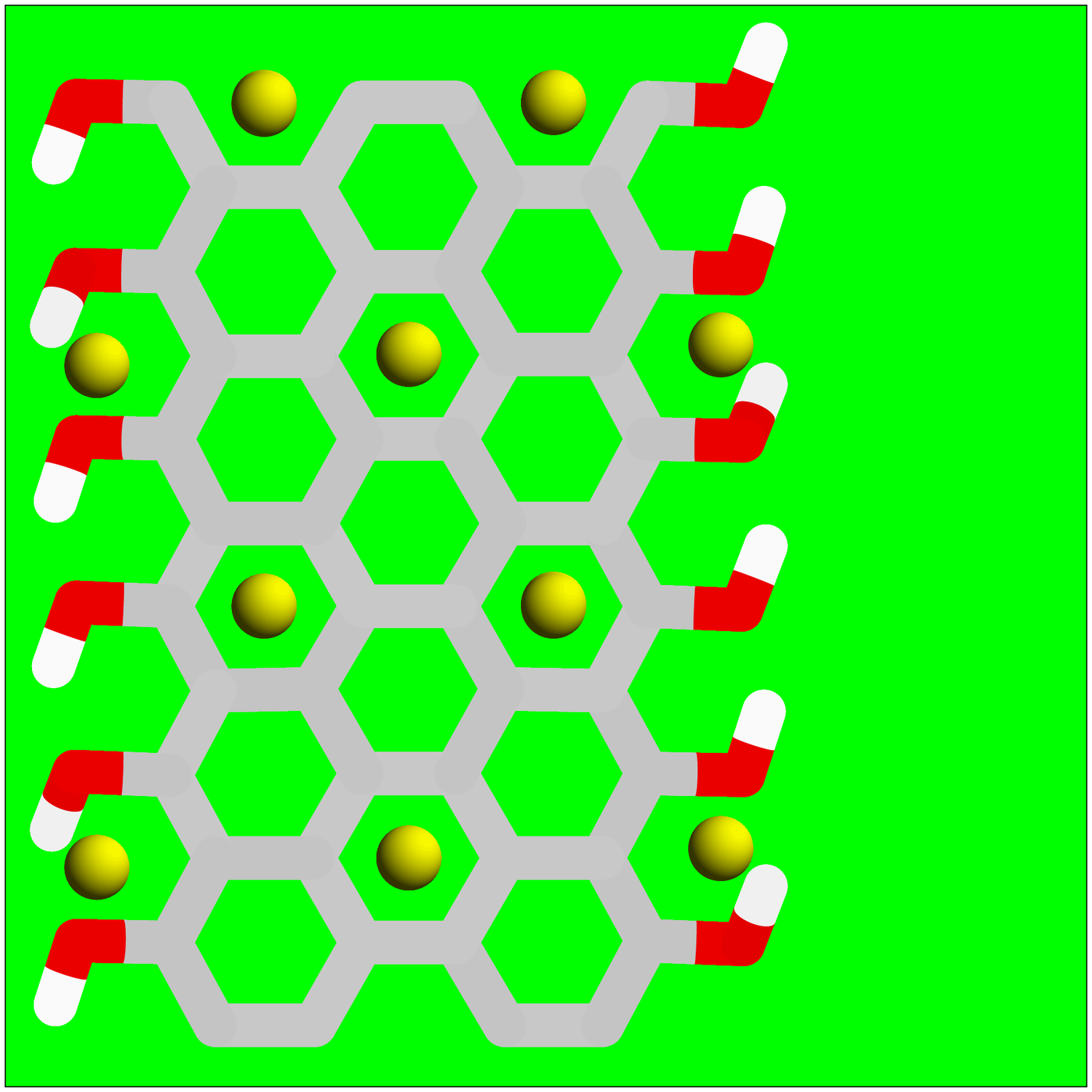}} (d)}
\caption[]
{\label{fig3} \noindent
%(a) Band offsets between isolated EC (red) and LiC$_6$ (blue).
(a) Basal plane LiC$_6$ plane (C$_{288}$Li$_{36}$);
(b)\&(c) C=O edge (C$_{192}$O$_{48}$Li$_{72}$), side and top views; 
in (b), only the outermost layer of Li chelated to 3 O~atoms each
are depicted;
(d) C-OH edge (C$_{192}$O$_{48}$H$_{48}$Li$_{32}$).
C-H edges (not shown) have no Li at edge sites.  The stick
figures depict graphite C, O, and H
atoms; color scheme as in Fig.~\ref{fig2}.
}
\end{figure}

Having re-examined breakdown in liquid EC, we turn to interfacial
reactions, the main thrust of this work.  Four layers of Li-intercalated
graphite are optimzed in a periodically replicated
29.74$\times$14.97$\times$15.06~\AA$^3$ cell with zig-zag
edges exposed (Fig.~\ref{fig3}).  The box size preserves the
density of the liquid region containing 32 EC after accounting for the
van der Waals radii of the electrode atoms.  Dangling $\sigma$-orbitals on
edge carbon atoms should be terminated with hydroxyl (C-OH), quinone
(C=O), carboxylic acid (COOH) functional groups, and/or
protons; only the first three are electrochemically active.\cite{mccreery}
We perform AIMD simulations on C=O, C-OH, and C-H terminated
LiC$_6$ interfaces, and use static calculations on the basal (0001)
plane (Fig.~\ref{fig3}a), impervious to Li$^+$, as a reference.  
Constant voltage conditions\cite{intro1,intro2} are not yet feasible
with AIMD.  Instead, our LiC$_6$ stochiometry\cite{holzwarth}
in the slab interior mimics a fully-charged battery anode.  To
some extent our simulation is akin to ``immersing'' SEI-free LiC$_6$,
pre-formed at low voltage, into EC liquid; this is analogous to the conditions
for SEI growth on Li metal, which yield EC-specific SEI similar to
those on graphitic anodes.\cite{intro1}

The optimal geometry when Li occupy all C=O edge sites is depicted
in Fig.~\ref{fig3}b and~\ref{fig3}c.  Each Li is chelated to two
and one O~atoms on successive graphite sheets (3 oxygens total),
with edge C=O bonds tilted out of the graphite planes to maximize
interaction with Li.   The mean Li-O distance is 1.98~\AA.
The Li chemical potential ($\mu_{\rm Li}$)
computed by randomly removing an edge Li atom is below -2.4~eV,
less than the $\mu_{\rm Li}$=-2.0~eV in the basal slab (Fig.~\ref{fig3}a),
suggesting the number of Li atoms is appropriate.  For C-OH edge LiC$_6$, 
when Li occupies every sixth edge site chelated with 2 O~atoms
on successive graphite sheets (4~total, Fig.~\ref{fig3}d),
$\mu_{\rm Li}$ is too {\it high} at -1.7~eV; too many Li
are present.  Nevertheless, this structure is considered
because mixed C=O/C-OH edges provide a proton source.  The
band offsets between these model electrodes and isolated EC molecules
are more favorable for electron transfer than that associated with
basal plane termination; this is shown in the ESI.

\begin{figure}
\centerline{(a) \hbox{\epsfxsize=1.50in \epsfbox{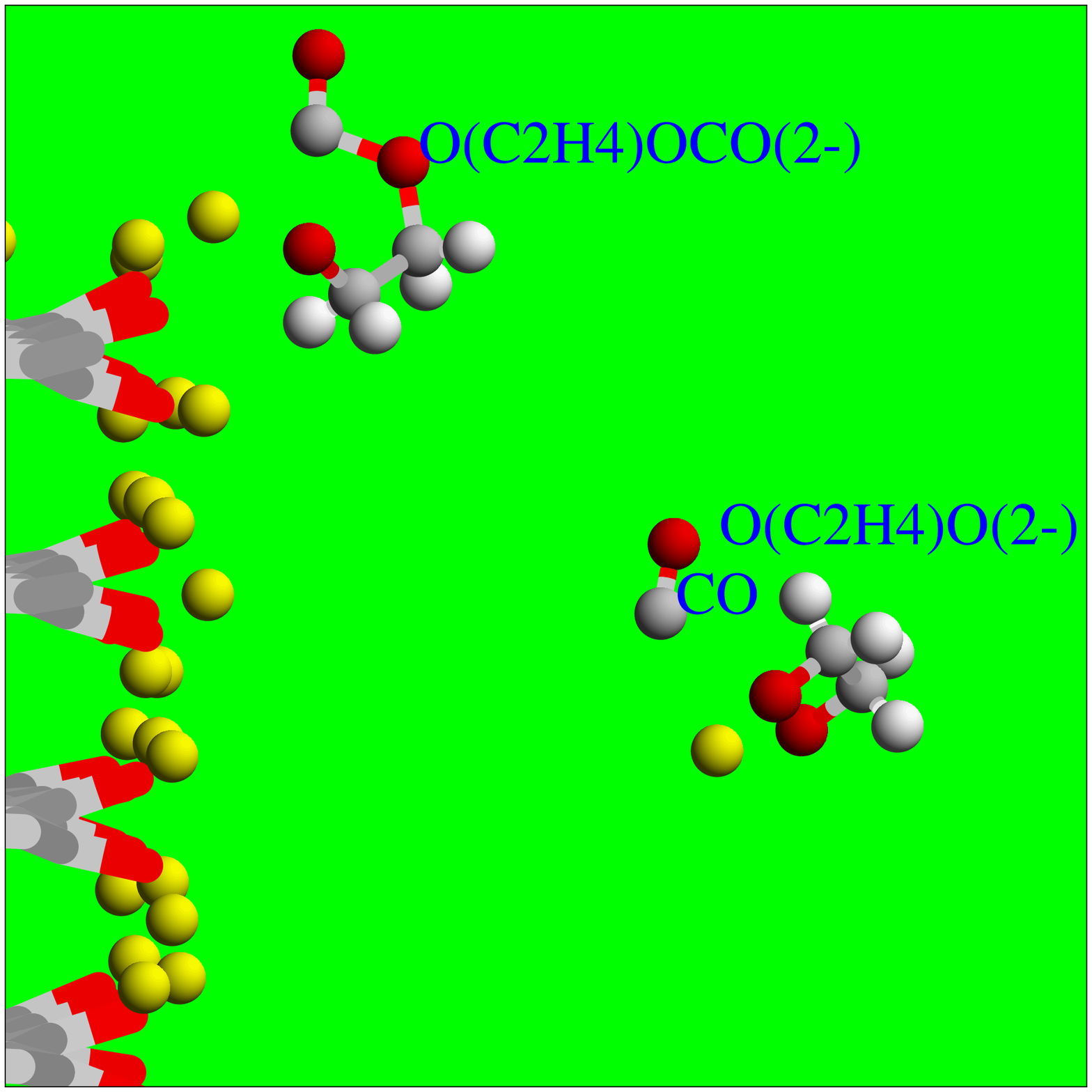}}
            \hbox{\epsfxsize=1.50in \epsfbox{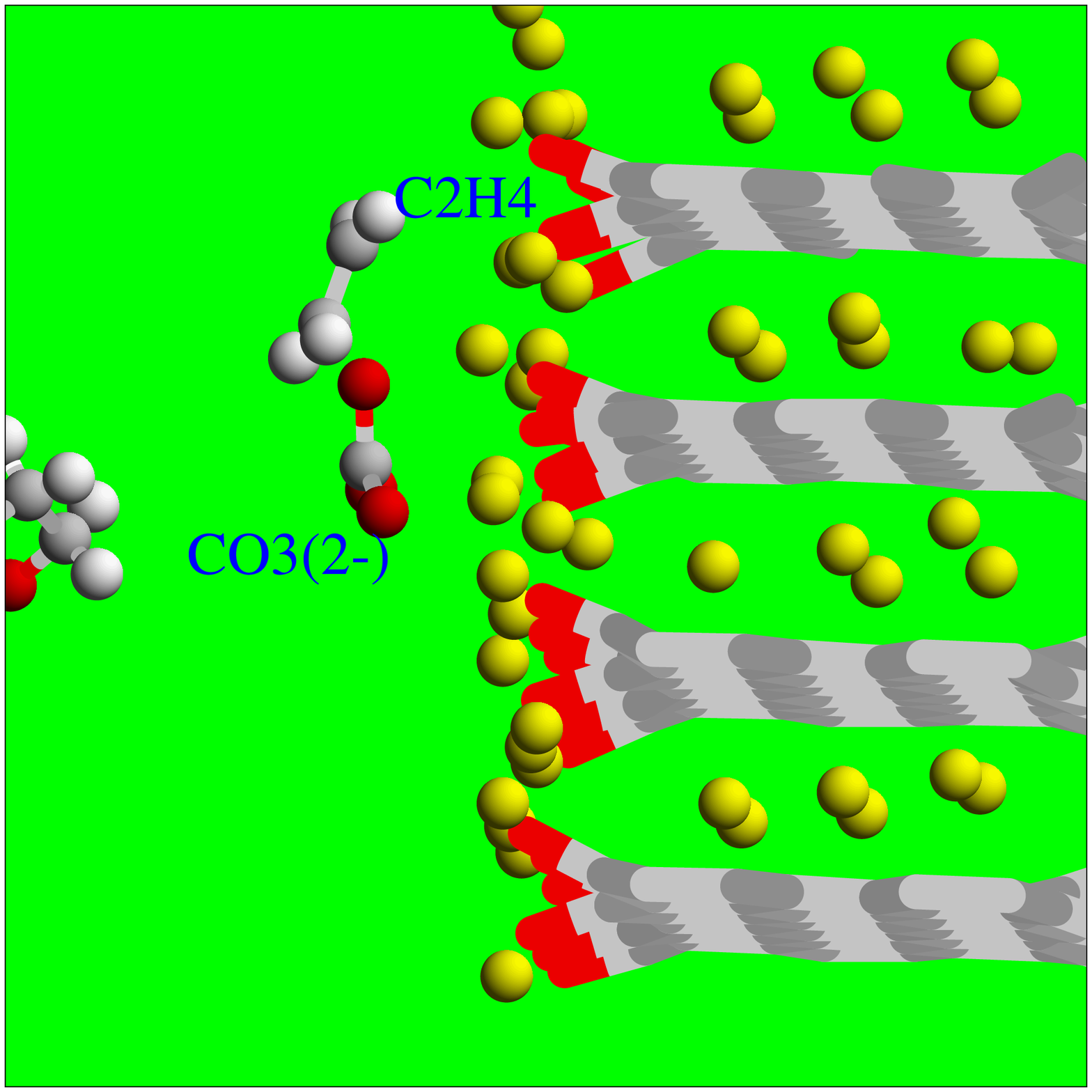}} (b)}
\centerline{(c) \hbox{\epsfxsize=1.50in \epsfbox{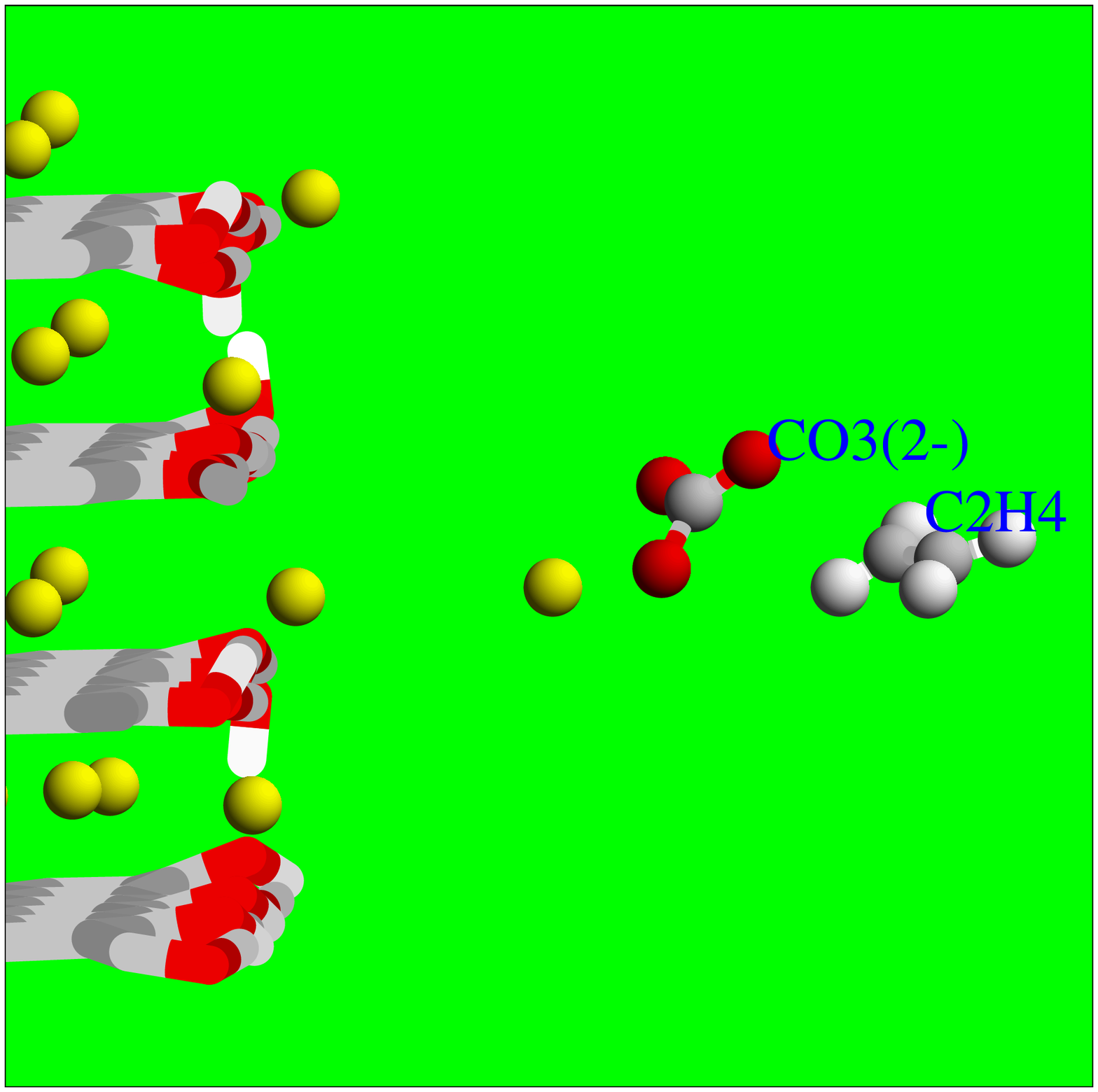}}
            \hbox{\epsfxsize=1.50in \epsfbox{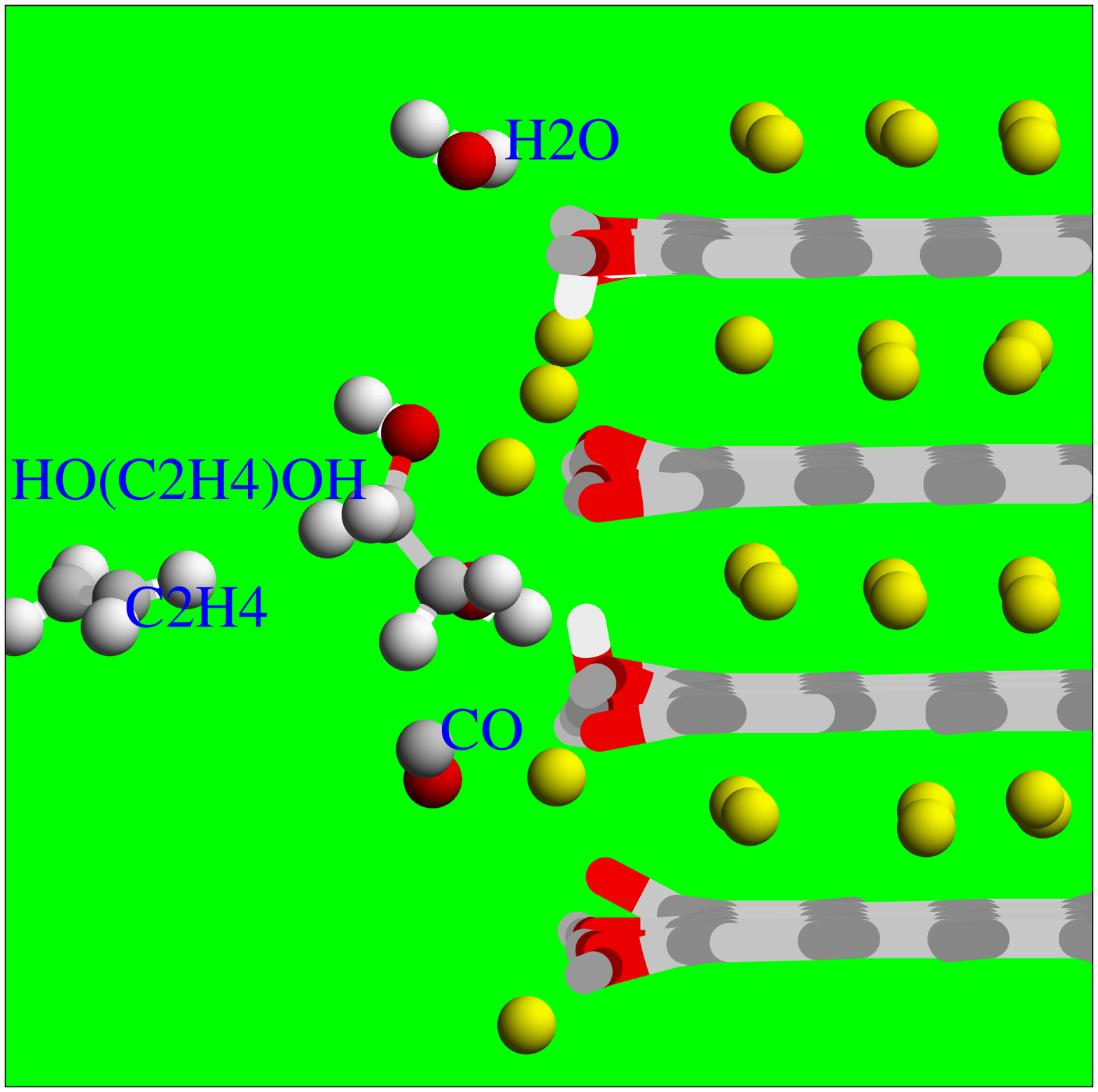}} (d)}
\caption[]
{\label{fig4} \noindent
EC breakdown products near the end of AIMD runs.  Intact EC 
(not shown) fill the empty spaces.
(a) C=O edge: a OC$_2$H$_4$OCO$^{2-}$, and
	a CO/OC$_2$H$_4$O$^{2-}$ pair (near a Li$^+$ away from surface).
(b) C=O (other surface): CO$_3^{2-}$/C$_2$H$_4$.
(c) C-OH edge: CO$_3^{2-}$/C$_2$H$_4$ pair near a Li$^+$
    away from surface.
(d) C-OH (other surface): CO/HOC$_2$H$_4$OH, and a H$_2$O
	from disproportionation of 3 C-OH.
Color scheme is as in Fig.~\ref{fig3}.
}
\end{figure}
 
AIMD simulations confine 32~EC between the C=O edge terminated LiC$_6$
surfaces (Fig.~\ref{fig3}c);\cite{init} in addition to the Li at the
C=O edges, an extra Li$^+$ resides in the liquid region.
Figures~\ref{fig4}a-b and Table~1 show that $e^-$ transfers from the
initially charge-neutral LiC$_6$ into EC to form a C$_2$H$_4$/CO$_3^{2-}$
pair (C$_{\rm E}$-O$_1$ cleavage), a CO/O(C$_2$H$_4$)O$^{2-}$ pair,
and a O(C$_2$H$_4$)OCO$^{2-}$ (C$_{\rm C}$-O$_1$ cleavages).  
The last moiety may ultimately crack into CO and O(C$_2$H$_4$)O$^{2-}$.  
The $-2|e|$ charge states are confirmed in the ESI.
In reactions occuring at electrode surfaces,
$e^-$ most likely flows directly to the decomposing EC's coordinated
to surface Li's without becoming first delocalized in the liquid.  As such,
this mechanism should not strongly depend on the EC liquid density or the
simulation cell size.  EC breakdown also occurs in the liquid
region where the dianionic products can bind to a Li$^+$.  In both
cases, the products are consistent with 2-$e^-$ addition to liquid
EC (Figs.~\ref{fig2}c \&~d), except
that CO formation has not occurred there yet.  
%As the reactions proceed, the net charge building up on our nanoscale
%model anode stops further $e^-$ transfer required for EC breakdown.  
Future work will consider counterions in the electric double layer, which
however equilibrate at timescales much slower than the observed reactions
and are neglected herein.

For the C-OH edge, both C$_2$H$_4$ + CO$_3^{2-}$ (Fig.~\ref{fig4}c)
and CO + O(C$_2$H$_4$)O$^{2-}$ (Fig.~\ref{fig4}d) products emerge
in picoseconds.  The O(C$_2$H$_4$)O$^{2-}$
then extracts H$^+$ from the electrode to form ethylene glycol, which
has been used as the chemical precursor (alongside carbonates) to
{\it synthesize} (CH$_2$CO$_3$Li)$_2$ outside battery settings.\cite{ethbicarb}
To our knowledge, CO products, observed in battery
charging experiments,\cite{yoshida,c13} have not been predicted in
calculations in the absence of explicit solvents.\cite{bal01,bal02,han,vollmer}
                                                                                
A spin-polarized triplet spin simulation yields qualitatively
similar conclusions (Table~1).  In contrast, no EC
decomposition is observed at the H-edge interface in 7~ps.
The edge dependences are consistent with their known electrochemical
activities,\cite{mccreery} but are seldom discussed in battery
studies because the initial carbon edges become masked by SEI formation.

\begin{table}\centering
\begin{tabular}{||l|c|c|c|c|c|c||} \hline
run & edge & spin & $t$ (ps)  & CO$_3^{2-}$ & OC$_2$H$_4$O$^{2-}$ &
 OCOC$_2$H$_4$O$^{2-}$ \\ \hline
1 & C=O & singlet & 7.0 & 1 & 1 & 1 \\
2 & C-OH & singlet & 7.0 & 1 & 1 & 0 \\
3 & C-OH & triplet & 6.8 & 2 & 0 & 1 \\
4 & C-H & singlet & 7.0 & 0 & 0 & 0 \\  \hline
\end{tabular}
\caption[]
{\label{table1} \noindent
EC decomposition products in interfacial simulations.  $t$ exclude
a 3.5~ps equilibration period during which an extra +4~$|e|$ charge
is imposed to hinder EC breakdown.  In Run~2, OC$_2$H$_4$O$^{2-}$
extracts two protons from C-OH groups to form ethylene glycol;
a OCHOC$_2$H$_4$OH is formed in Run~3.}
\end{table}
 
Our novel predictions suggest the following mechanism for initial SEI
growth.  Fast EC decomposition initiates at graphite edge regions rich in
oxidized sites.\cite{ein}  Both C$_{\rm E}$-O$_1$ and C$_{\rm C}$-O$_1$
cleavages occur, yielding CO$_3^{2-}$ and OC$_2$H$_4$O$^{2-}$ respectively. 
OC$_2$H$_4$O$^{2-}$, or HOC$_2$H$_4$OH if a proton source exists,
potentially reacts with CO$_2$ or CO$_3^{2-}$ at longer time scales to form
(CH$_2$CO$_3$Li)$_2$,\cite{ethbicarb} a main SEI component.  
C$_{\rm C}$-O$_1$ cleavage and OC$_2$H$_4$O$^{2-}$ products
(Fig.~\ref{fig1}a) are faciliated by Li$^+$ and other ionic products
coordinating to carbon edges.  However, slower $e^-$ transfer to the
solvent may mandate different decomposition mechanisms
at later stages of SEI growth.\cite{review,intro1,intro2}

In conclusion, AIMD simulations can yield new insights
concerning ethylene carbonate decomposition in electron-rich anode
regions.  With explicit treatment of the EC liquid environment, an excess
electron can induce EC decomposition within AIMD timescales.  On pristine
graphitic electrodes, carbon edge terminations strongly affect
EC breakdown.  C=O edges give strong driving forces and support
two EC decomposition pathways, yielding both C$_2$H$_4$ and CO gas
and ionic products.  CO evolution, observed in experiments,\cite{c13}
is predicted for the first time.
Although our reported time scales may be somewhat affected by the DFT
method used, the reactions pertinent to initial stages of SEI growth
are surprisingly fast, which potentially
opens the way for the versatile AIMD method to investigate the
decomposition mechanisms of other solvent or additive molecules.\cite{proofs}
 
We thank John Sullivan, David Ingersoll, Kevin Zavadil, Kang Xu, and Perla
Balbuena for useful discussions.  This work was supported by the Department
of Energy under Contract DE-AC04-94AL85000.  Sandia is a multiprogram
laboratory operated by Sandia Corporation, a Lockheed Martin Company,
for the U.S.~Deparment of Energy.  KL was partly supported by
Nanostructures for Electrical Energy Storage,
an Energy Frontier Research Center funded by the U.S. Department of Energy,
Office of Science, Office of Basic Energy Sciences under Award
Number DESC0001160.  The electronic supporting information is available
at {\tt http://www.rsc.org}.


\begin{references}
\bibitem{init}
AIMD trajectories are initialized using equilibrated classical force
fields\cite{li_balbuena} generated configurations.  When pre-equilibrating
EC-LiC$_6$ interfaces, all atoms in the solid are frozen.  During
AIMD interfacial runs, the C, O, and H atoms at the graphite edge and
all Li atoms can move.  The costliest, 7~ps spin-polarized interfacial
trajectory took 10$^5$ cpu-hours on Intel processors.
\bibitem{proofs}
Notes added in proofs: preliminary AIMD/PBE0 simulations yield
a lower tendency towards two-electron induced CO formation in
EC liquid (Fig.~\ref{fig2}d),\cite{new} but show that the pertinent C-O
bond breaking readily occurs at the electrolyte-electrode interface.
\end{references}
\end{document}